\documentclass[prb,aps,floats,showpacs,tightenlines, 
	twocolumn,floatfix,letterpaper]{revtex4}
\usepackage{graphicx,color,times}

\newcommand{\be}{\begin{equation}}
\newcommand{\ee}{\end{equation}}
\newcommand{\bea}{\begin{eqnarray}}
\newcommand{\eea}{\end{eqnarray}}
\newcommand{\nn}{\nonumber}

\newcommand{\rr}{{\bf r}}
\newcommand{\qq}{{\bf q}}
\newcommand{\kk}{{\bf k}}

\newcommand{\bwt}{\begin{widxetext}}
\newcommand{\ewt}{\end{widetext}}
\newcommand{\vvv}{\vspace{-.00cm}}

\begin{document}

\title{Collective Excitations in Quantum Hall Liquid Crystals:
Single-Mode Approximation Calculations}

\author{Cintia M.\ Lapilli}
\affiliation{
	Department of Physics and Astronomy,
     University of Missouri--Columbia,
     Columbia, Missouri 65211, USA}  

\author{Carlos Wexler}
\affiliation{
	Department of Physics and Astronomy,
     University of Missouri--Columbia,
     Columbia, Missouri 65211, USA}  

\date{August 19, 2005, new version: December 30, 2005}

\begin{abstract}
A variety of recent experiments probing the low-temperature transport
properties of quantum Hall systems have suggested an interpretation in
terms of liquid crystalline mesophases dubbed {\em quantum Hall liquid
crystals}.  The single mode approximation (SMA) has been a useful tool 
for the determination of the excitation spectra of various
systems such as phonons in $^4$He and in the fractional quantum Hall
effect. In this paper we calculate (via the SMA) the spectrum of
collective excitations in a quantum Hall liquid crystal by considering
{\em nematic}, {\em tetratic}, and {\em hexatic} generalizations of
Laughlin's trial wave function having two-, four- and six-fold broken
rotational symmetry, respectively.  In the limit of zero wavevector
$\qq$ the dispersion of these modes is singular, with a gap that is
dependent on the direction along which $\qq=0$ is approached for {\em
nematic} and {\em tetratic} liquid crystalline states, but remains
regular in the {\em hexatic} state, as permitted by the fourth order
wavevector dependence of the (projected) oscillator strength and
static structure factor. 
\end{abstract}

	
\maketitle

\section{Introduction}
\label{sec:introduction}

For more than two decades two-dimensional electron systems (2DES), and in
particular the quantum Hall effect (QHE)
\cite{qhe,laughlin83,perspectives} have 
been constant sources of complex and unexpected behavior,
perhaps with no equal in the realms of condensed matter
physics.   The unique combination of extremely high
mobilities ($\mu \sim 10^7$ m/Vs) in GaAs/Al$_x$Ga$_{1-x}$As
heterostructures, low temperatures ($T < 100$ mk), enhancement of
interactions due to the reduced dimensionality, and relative
quenching of the kinetic energy in strong magnetic fields due 
to Landau level (LL) quantization, has allowed the emergence of complex
and striking behavior due to subtle correlation effects.  In fact, from
the discovery of the integer and fractional QHE's in the early eighties
\cite{qhe} (leading to the award of two Nobel prizes),
to the existence of novel fractionally charged \cite{laughlin83} and
composite particles \cite{cf00} and many newer interesting 
many-body phenomena, the QHE has been a consistently active and exciting
area of research.  

The physics of 2DES in partially filled LL's is inherently complex due
to the high degeneracy of the ``unperturbed'' ground state (i.e.\
without the Coulomb interaction).  The first successful theoretical
approach to this system was proposed shortly after the discovery of
the FQHE \cite{qhe} by Laughlin, who proposed his famous
trial-wavefunction \cite{laughlin83}
\vvv
\be
\label{eq:laughlin}
\Psi_{1/m}= \prod_{i<j}^{N} \, (z_i-z_j)^{m} \,
         e^{-\frac{1}{4}\sum_{k=1}^{N} {|z_k|^2} } \,,
\vvv
\ee
to describe states at filling factor $\nu = 1/m$, with $m$ an odd
integer.  Here $z_j = x_j + i y_j$ is the position of the $j^{\rm th}$
electron in the complex plane and we work in units of the magnetic
length $l_0 = [\hbar/eB]^{1/2}$.   The ``goodness'' of Laughlin state
originates in that the nodal hyper-surfaces on which the 
many-body wavefunction vanishes coincide with the ones in which the
particles are in contact.  Whereas the vanishing of the wavefunction
when two (spin aligned) electrons are in contact is required by
Fermi statistics, Laughlin  state has multiple nodes at those points,
thus reducing the Coulomb repulsion.

Later, Jain proposed \cite{cf00} the elegant conceptual framework
of the composite fermions (CF) which unifies all
hierarchies of integer and fractional QHE's, along with 
the intermediate states in between QH plateaus:
an even number $2p$ of vortices is attached to each electron,
also lowering the Coulomb energy, 
and (in a ``mean field'' approximation) the ``magnetic fluxes''
associated with the vortices  lead to a reduction of the {\em
effective} magnetic field: $B^* = B - 2 p \Phi_0 n_e$ ($\Phi_0 = h/e$
is the flux quantum and $n_e$ is the electron density),
resulting in an {\em effective}
filling factor $\nu^* = \nu/(1 - 2 p \nu)$.  It is easy to see that
for the strongest FQHE states $\nu^*$ is an integer (e.g. for $\nu = 1/3$
and $p=1$ $\nu^* = 1$), leading to the interpretation of the FQHE of
electrons as a simple IQHE of the CF's; whereas the intermediate
regions (e.g. $\nu = 1/2, 1/4$) correspond to $\nu^* \rightarrow
\infty$, i.e.\ a vanishing effective magnetic field: $B^* = 0$.

Since 1999, magnetotransport experiments have uncovered a variety of
surprising results at low temperatures ($T \lesssim 100$ mK), for
example:  extreme anisotropies \cite{lilly99andothers99} and apparent 
competition between different ordered phases \cite{competition} in the
intermediate regions between quantum Hall plateaus at high LL's, the
melting transition between a Wigner crystal (WC) at $\nu \simeq
1/7$,\cite{melting17} and the microwave conductivity  
evidence of structural phase transitions in partially filled
LL's.\cite{lewis04}   A large body of evidence corresponding 
to seemingly distinct phenomena may be partly understood in
terms of a simple conceptual point: that the many-body states involved
have an intrinsic crystalline or liquid crystalline
order,\cite{foglermoessner,fradkin99,wexlerKT,brs,liqcryst04,many,
radzihovsky02fogler04,oursother,elasticity}  be it {\em
smectic},\cite{foglermoessner,wexlerKT}  {\em
nematic},\cite{fradkin99,wexlerKT,brs,liqcryst04} {\em tetratic} or 
{\em hexatic}.\cite{liqcryst04}

Generalizations of Laughlin wavefunction [Eq.\ (\ref{eq:laughlin})]
with discrete broken rotational symmetry (BRS) have been proposed in the
past \cite{brs,liqcryst04,joynt96balents96} as candidates for
nematic or hexatic states \cite{brs,liqcryst04}
in order to understand anisotropic transport observed in the
intermediate regions,\cite{lilly99andothers99} or the melting of the WC
at $\nu = 1/7$.\cite{melting17}
In fact, the motivation for these states arises from the fact that 
{\em it is generally expected} that melting in 2D may occur through
a topological Koszterlitz-Thouless-type (KT)
transition.\cite{kosterlitz73}  For 2D melting, the
reliable Kosterlitz-Thouless-Halperin-Nelson-Young (KTHNY) theory
\cite{kosterlitz73,hny} predicts that, in fact, an
{\em intermediate liquid crystalline phase} will exist between a solid
and a liquid phase, which will exhibit no {\em translational} 
order, and only a quasi-long-range order for the orientational order
below the KT transition temperature.  These arguments were used by
Wexler and Dorsey \cite{wexlerKT} to calculate qualitatively correct 
anisotropic-isotropic transition temperatures for the quantum Hall
liquid crystal in the transitional regions at high
LL's.\cite{lilly99andothers99}


In this paper we consider the spectrum of collective excitations for a
family of liquid crystalline states in a partially filled LL.
These states are generated so as to
satisfy the following criteria which we consider reasonable for
understanding the dynamics:  
(i) states must obey Fermi statistics, i.e.,
they must be {\em odd} under the exchange of any pair of electrons;
(ii) the states must be translationally invariant (far enough from 
boundaries); 
(iii) there must be a broken rotational symmetry belonging to the
proper symmetry group (i.e., $C_2$ for a {\em nematic}, $C_4$ for a
{\em tetratic}, and $C_6$ for a {\em hexatic}; additional symmetries
are possible in principle, e.g. with a $C_{10}$ symmetry, we have not
explored such possibilities); (iv) states and excitations must reside
entirely in the LLL to avoid the large cyclotron energy cost $\hbar
\omega_c$. Note, as we will show later, that the properties of {\em
any} excited LL may be readily obtained from the properties of the LLL. 

\begin{figure}
\begin{center}
\leavevmode
\includegraphics[width=3.3in]{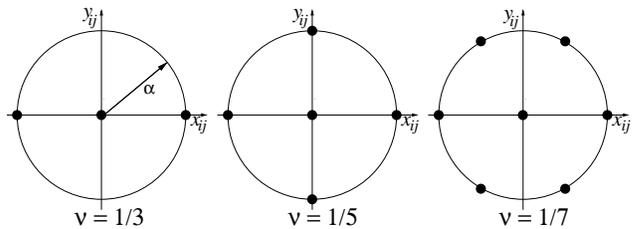}
\vspace{-.5cm}
\end{center}
\caption{\label{fig:brs} 
	Nodal distribution for $z_{ij} \equiv z_i - z_j$ for a quantum
	Hall {\em nematic} at $\nu = 1/3$, {\em tetratic} at $\nu =
	1/5$, and {\em hexatic} at $\nu = 1/7$.
} 
\vspace{-.5cm}
\end{figure}

States that satisfy the abovementioned requirements have been proposed
and studied in detail \cite{brs,liqcryst04,joynt96balents96}
for filling factors $\nu = 1/3$, 1/5, and 1/7.
These are found by splitting the ``extra'' vortices of the Laughlin
(or other CF) states around the electron, while obeying the required
symmetries: 
 %
\vvv
\bea
\label{eq:brs}
\Psi_{1/(2p+1)}^\alpha &=& 
         \left\{ \prod_{i<j}^{N}
        \left[ \prod_{\mu=1}^{2p} (z_i-z_j-\alpha_\mu) \right]
    \right\}  \times
\nn \\
 &&\times     \prod_{i<j}^{N} \, (z_i-z_j) \,
        e^{-\frac{1}{4}\sum_{k=1}^{N} |z_k|^2} 
\,,
\vvv 
\eea
\noindent
where the complex numbers $\alpha_\mu$ are distributed in pairs of
opposite value in the complex plane (to satisfy Fermi statistics).
For the states with the highest discrete symmetry at each filling
factor we may take 
\be
\alpha_\mu = \alpha \ e^{i \, 2\pi (\mu-1)/2p} \,,
\ \ \ \mu \in \{1,2,\ldots, 2p\}\,,
\ee
and without loss of generality $\alpha$ can be taken to be real. The
wavefunction in Eq.\ (\ref{eq:brs}) represents a homogeneous liquid
crystalline state at $\nu=1/(2p+1)$, is
anti-symmetric, lies entirely in the LLL, and is smoothly connected to
the isotropic Laughlin state for $\alpha = 0$.  
Figure \ref{fig:brs} depicts the nodal distribution for various states
of Eq.\ (\ref{eq:brs}). A remarkable feature
of these states is that they posses an underlying charge density
wave (CDW), but these CDWs are melted
by fluctuations, and overall the system is translationally
invariant.\cite{liqcryst04}

\section{The Single Mode Approximation}

To calculate the excitation spectrum we use the single mode
approximation (SMA),\cite{feynman,gmp-sma}  which reliably provides
the first moment (mean) of the energy  of the excitations (for a given
wavevector $\kk$) that are coupled to the ground state by means of the
density operator.\cite{feynman,gmp-sma,park00}  The SMA was 
first used by Feynman in 1953 to accurately calculate the spectrum of
phonons in superfluid $^4$He.\cite{feynman}  The essence of the method
originates on the assumption  that the ground state of a system of
bosons has a scarcity of long-wavelength excitations.  Under
those circumstances, the variational wavefunction for an excitation
corresponding to a density-wave can be written as
\vvv
\be
\label{eq:HePhi}
\phi_{\bf k}({\bf r}_1, \ldots, {\bf r}_N) = N^{-1/2} \rho_{\bf k} 
\psi_0({\bf r}_1, \ldots, {\bf r}_N) 
\,, 
\vvv
\ee
where $\rho_{\bf k} = \sum_{j=1}^N e^{-i {\bf k}\cdot{\bf r}_j}$ is
the density operator, and $\psi_0$ is the many-body ground state
(which is, in fact, unknown for $^4$He).  Note that this trial state
automatically builds in the favorable correlations of the ground
state. The energy of this excited state, $\Delta({\bf k}) = \langle
\phi_{\bf k} | H - E_0 | \phi_{\bf k} \rangle / \langle \phi_{\bf k} |
\phi_{\bf k} \rangle$, can be simply evaluated: 
\vvv 
\be
\label{eq:deltaHe}
\Delta({\bf k}) 
= \frac{N^{-1}\langle \psi_0 |  \rho_{\bf k}^{\dag} [H, \rho_{\bf k}]|\psi_0 
\rangle} 
{N^{-1}\langle \psi_0 |  \rho_{\bf k}^{\dag}\rho_{\bf k}|\psi_0 \rangle} 
\equiv
\frac{f({\bf k})}{S({\bf k})}
\,,
\vvv
\ee
In the last term the numerator is the ``oscillator strength'' and takes on
the universal value $f({\bf k}) = \hbar^2 k^2/2m$, and $S({\bf k})$ is
the static structure factor, which is directly measurable by means of
neutron scattering (it is here, in fact, where the He-He correlations
of the ground state are ``encoded'').  Using the experimental results
for $S({\bf k})$, Feynman could calculate
a spectrum of remarkable quality, showing the phonon-like spectrum at
small wavevectors, and a roton minimum at wave-vectors comparable
to the inverse interatomic distance.\cite{feynman}

\section{The Single Mode Approximation in the Quantum Hall Effect}
\label{sec:gmp-sma}

\begin{figure*}
\begin{center}
\leavevmode
\includegraphics[width=6.in]{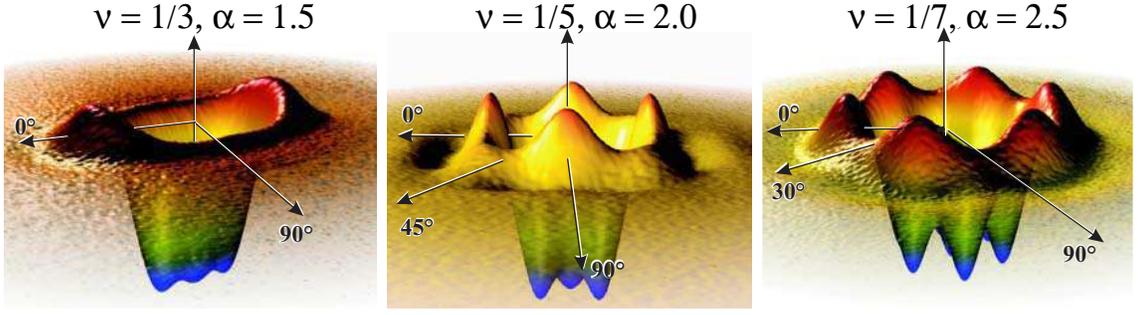}
\vspace{-.5cm}
\end{center}
\caption{\label{fig:gxys} 
        (Color online) Pair correlation function $g({\bf r})$ for 
        a $\nu = 1/3$ {\it nematic}, a $\nu = 1/5$ {\it tetratic}, 
        and a $\nu = 1/7$ {\it hexatic}.  In all cases 
        $g({\bf r}) = {\cal O}[r^{2/\nu}]$ for small $r$ and 
        $g(\infty) \rightarrow 1$. The fact that $g(\rr) \simeq 1$
	over a large area is a guarantee that the calculations are
	able to reproduce the thermodynamic limit.}
\vspace{-.5cm}
\end{figure*}

The applicability of the SMA to {\em fermion} systems is also well
established for two- and three-dimensional systems in the absence of
magnetic fields, giving a good approximation for the plasmons at long
wavelength. For 2DES in presence of a magnetic field it correctly gives 
the zero-wavevector magnetoplasmon at $\omega_c = eB/m_e$, a result that
is guaranteed by Kohn's theorem,\cite{kohn61}
which states that the dipolar excitation is saturated by the cyclotron
mode (this results in the modes of interest---the {\it intra-LL}
excitations---having {\it quadrupolar} character, i.e. with an
oscillator strength $\bar{f} \propto q^4$).  

For excitations fully contained within a single LL, the cyclotron mode
is not of primary interest.  In 1985  Girvin, MacDonald an
Platzman (GMP) proposed an ingenious ansatz for {\it projected} excited
states:\cite{gmp-sma}
\vvv
\be
\label{eq:gmp}
|\psi_{\bf q}\rangle = \bar{\rho}_{\bf q} \, |\psi_0\rangle
\,, 
\vvv
\ee
where $\bar{\rho}_{\bf q}$ is the {\it projected} density
operator:\cite{girvin84}
\vvv
\bea
\label{eq:rhop}
\bar{\rho}_{\bf q} &=& \sum_{m,m'} \langle 0, m'| e^{-i {\bf q} \cdot {\bf r}}
	| 0, m \rangle a^{\dag}_{0,m'} a_{0,m} 
\\
	&=& \sum_{j=1}^N \overline{e^{-i {\bf q} \cdot {\bf r}}} 
	= \sum_{j=1}^N e^{-|q|^2 /2} \, e^{-i q^* z_j/2} 
		\, e^{-i q \frac{\partial}{\partial z_j}} \nn
\,,
\vvv
\eea
where $| 0, m \rangle$ correspond to single-particle states in the lowest
LL and angular momentum $m$, and $a^{\dag}_{0,m}$ is the particle creator
operator for such state.  As in Feynman's ansatz [Eq.\
(\ref{eq:HePhi})], Eq.\ (\ref{eq:gmp}) preserves the favorable
correlations of the ground state.  The exclusion of inter-LL
excitations eliminates the problem with the saturation of the dipolar
mode.  
The excited states have a compelling description, in first quantized
form:\cite{girvin84} 
\bea
\bar{\rho}_{\bf q} \, \psi(z_1,\ldots,z_N) &=&
	\sum_{j=1}^N e^{-|q|^2 /2} \, e^{-i q^* z_j/2} \,\times \\ 
	&& \hspace{-1cm}
	\times \, \psi(z_1,\ldots,z_{j-1},z_j-iq,z_{j+1},\ldots,z_N)
\nn \,, 
\eea
which corresponds to shifting each electron by $\hat{e}_z \times {\bf q}$
and superimposing these $N$ configurations with an amplitude $e^{-i q^*
z_j/2}$.

Similarly to Eq.\ (\ref{eq:deltaHe}), the excitation spectrum
can be readily obtained
\vvv
\be
\bar{\Delta}_{\bf q} 
	=\frac{(2N)^{-1}\langle \psi_0 | 
		[\bar{\rho}_{\bf q}^{\dag},[\bar{H},\bar{\rho}_{\bf q}]]
	| \psi_0 \rangle}
	{{N}^{-1}\langle \psi_0 | 
		\bar{\rho}_{\bf q}^{\dag} \bar{\rho}_{\bf q}
	| \psi_0 \rangle }
	\nn \\
\equiv
\frac{\bar{f}({\bf q})}{\bar{S}({\bf q})}
\,.
\vvv
\ee
The projected oscillator strength comes from the
non-commutation of the projected density operator with terms in the
{\em potential} energy part of the Hamiltonian also projected
onto the LLL:
\vvv
\be
\bar{H} 
	= \frac{1}{2}\sum_{\bf q} v_{\bf q} ( \bar{\rho}_{\bf q}^{\dag} 
	\bar{\rho}_{\bf q} - Ne^{-q^2 l^2/2})
\,.
\vvv
\ee
Since $[\bar{\rho}_{{\bf k}}, \bar{\rho}_{\bf q}] =
(e^{k^*q/2}-e^{kq^*/2})\bar{\rho}_{{\bf k}+{\bf q}}$,
we find \cite{gmp-sma}  
\bea
\label{eq:fq}
\bar{f}({\bf q}) &=& 2 \, e^{-|q|^2/2} \sum_{\bf k}
	\sin^2(\frac{{\bf q} \times {\bf k}}{2})
	\bar{S}({\bf k}) \times \\
&&  \hspace{2.5cm}\times
	[v_{{\bf k}-{\bf q}} e^{{\bf k} \cdot {\bf q}} 
	e^{-|q|^2/2} - v_{\bf k}] 
\,, \nn
\vvv
\eea

For its part, the projected static structure factor $\bar{S}({\bf q})$
can be calculated from:\cite{gmp-sma,girvin84}
\vvv
\be
{\bar{S}}({\bf q}) = S({\bf q}) - (1 - e^{-q^2/2})
\,,
\vvv
\ee
where $S(\qq)$ is the {\em unprojected} static structure factor:
\vvv
\be
S(\qq) - 1 = \rho_0 \int d^2r \, e^{-i \qq \cdot \rr} \,
	[g(\rr) - 1] \,,
\vvv
\ee
the Fourier transform of the pair correlation function
\vvv
\be
g({\bf r}) = {n_e^{-2}}
\langle \sum_{i\neq j}^N \delta({\bf r}_i - {\bf r})\delta({\bf r}_j )
\rangle
\vvv
\ee
which is obtainable from the ground state via, e.g., Monte Carlo (MC)
simulations.\cite{gmp-sma,brs,liqcryst04}  In our case, we considered
BRS states [Eq.\ (\ref{eq:brs})] corresponding to a $\nu = 1/3$ {\it
nematic}, a $\nu = 1/5$ {\it tetratic}, and a $\nu = 1/7$ {\it
hexatic}.  Figure \ref{fig:gxys} depicts the pair correlation
function for various states.  
In all cases, correlation functions and SMA excitation
spectra were computed for numerous $\alpha$'s.  
The angle dependence significantly increases the burden in the MC
simulations since the full angle-dependent $g({\bf r})$ is needed,
rather than the considerably  simpler angle-averaged $g(r)$ of the
isotropic systems.  The accurate calculation of $\bar{f}(\qq)$, with
its angular-dependent exponentially large factors required high quality
$\bar{S}(\qq)$ and hence $g(\rr)$.  To put things in perspective, to
achieve ca.\ 1\% accuracy in $g(\rr)$, ${\cal O}[10,000]$ counts were
accumulated for each $0.01\times 0.01$ (in units of $l_0$) $\rr$-box in
the original histograms (the small boxes were necessary to have 
precision in the Fourier transform at high wavevectors).  This
required, for each filling factor $\nu$ and anisotropy generating
$\alpha$, to run approximately  4--8$\times 10^7$ MC steps in systems
of $N_e=$ 200--400 electrons taking 100--200 cpu$\times$days of computation
for each $(\nu,\alpha)$ in a 2 GHz Athlon cluster (see
Ref. [\onlinecite{liqcryst04}] for further details).  A relatively
large $N_e$ is required so that the simulations are able to reproduce
a system in the thermodynamic limit.  The fact that $g(\rr) \simeq 1$
over a large area is a guarantee of that achievement (Fig.\
\ref{fig:gxys}). 

\begin{figure*}
\begin{center}
\leavevmode
\includegraphics[width=6in]{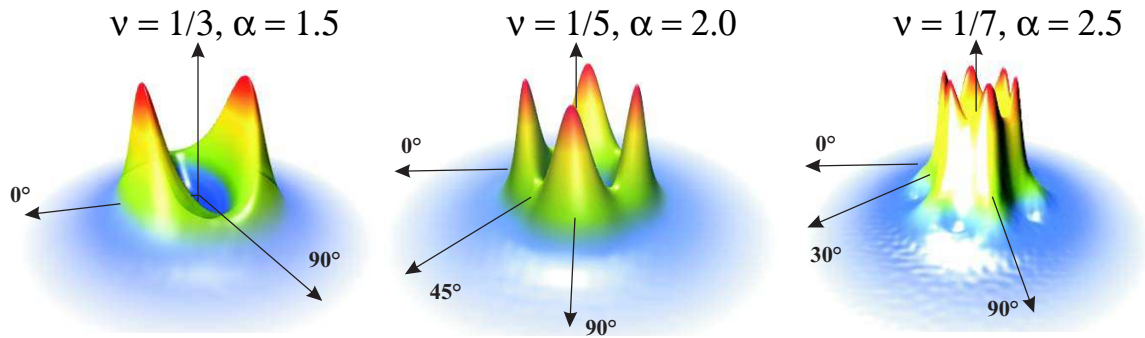}
\vspace{-.5cm}
\end{center}
\caption{\label{fig:sqprojs} 
        (Color online) {\it Projected} static structure factors
	$\bar{S}({\bf q})$ for a $\nu = 1/3$ {\it nematic}, a $\nu =
	1/5$ {\it tetratic}, and a $\nu = 1/7$ {\it hexatic}. In all
	cases, $\bar{S}({\bf q}) = {\cal O}[q^4]$ for small $q$, and 
        $\bar{S}(\infty) \rightarrow 0$.}
\vspace{-.5cm}
\end{figure*}

Figure \ref{fig:sqprojs} shows the {\em projected} static structure
factors $\bar{S}({\bf q})$ for a $\nu = 1/3$ {\it nematic}, a $\nu =
1/5$ {\it tetratic},s and a $\nu = 1/7$ {\it hexatic}.  
From $\bar{S}({\bf q})$, the oscillator strength
$\bar{f}({\bf q})$ is computed using Eq.\ (\ref{eq:fq}).  
Analysis of $\bar{S}({\bf q})$ and $\bar{f}({\bf q})$ shows that {\it
both} are ${\cal O}[q^4]$ for small ${\bf q}$.\cite{gmp-sma}
This restriction on the the small wavevector behavior originates in
Kohn's theorem,\cite{kohn61} as all of the ${\cal O}[q^2]$ pieces in
the {\it unprojected} parts are saturated by the (uninteresting)
inter-LL excitations at $\hbar \omega_c$.  

\begin{figure*}
\begin{center}
\leavevmode
\includegraphics[width=6in]{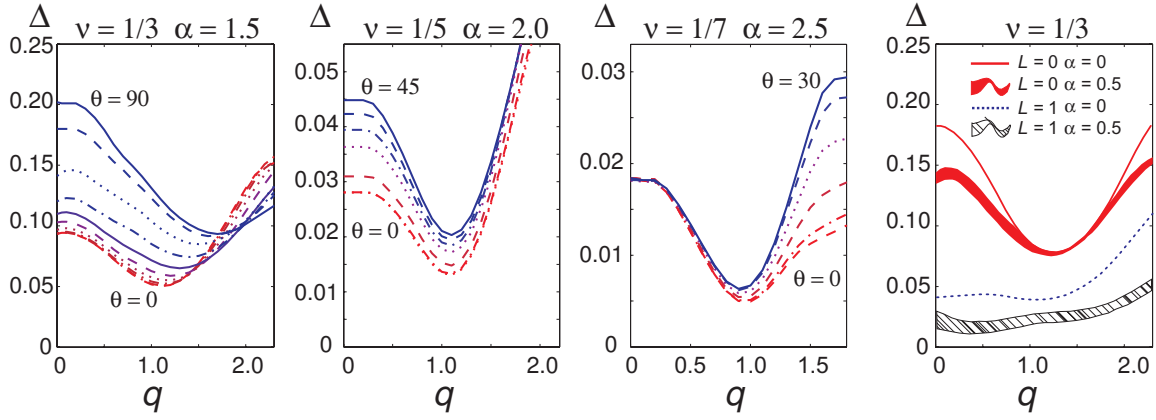}
\vspace{-.5cm}
\end{center}
\caption{\label{fig:spectra} 
        (Color online) Single mode approximation spectra for a $\nu = 1/3$ 
        {\it nematic}, a $\nu = 1/5$ {\it tetratic}, and a $\nu = 1/7$
	{\it hexatic}. Note the dramatic angular dependence of the
	spectra, and the appearance of a singular gap as $q
	\rightarrow 0$ for the nematic and tetratic.  The lower right
	panel shows also the spectum for the $\nu = 1/3$ 
        {\it nematic} in the first excited LL ($L=1$).
} 
\vspace{-.5cm}
\end{figure*}

Figure \ref{fig:spectra} presents some of our results for
the excitation spectra in the lowest LL.  The results are consistent
with those obtained by GMP for the isotropic $\nu = 1/3$ and 1/5 FQHE
cases \cite{gmp-sma} which were qualitatively confirmed
experimentally.\cite{pinczuk}  They show that the collective
excitation spectrum 
remains gaped, albeit with a deep magnetoroton, for modes coupled to
the ground state via the density operator.  Not surprisingly, BRS
states have significant anisotropy in their spectra.  However, it is
noteworthy that for the nematic and tetratic cases the spectrum is
{\em singular}, with an angle dependence on the excitation energy
$\Delta(\qq)$ as $\qq \rightarrow 0$.   By contrast, the hexatic
liquid crystal has a regular spectrum in the long wavelength limit.
The apparent disparity have, of course, to do with the different
rotational symmetries of the different states:   
as $\bar{\Delta} = \bar{f}/\bar{S}$, and both numerator and denominator
are ${\cal O}[q^4]$ for small ${\bf q}$, there is no possibility of
generating a $C_6$ symmetric form with terms that can only depend on
$q_x^4$, $q_y^4$, and $q_x^2 q_y^2$.  This can also be understood from
the point of view of an effective elasticity theory for
2DES (valid in the long wavelength limit, $q \rightarrow
0$):\cite{vignale98} the elasticity tensor being of the 
fourth rank is not compatible with a six-fold rotational symmetry.  This 
elastic interpretation will be published elsewhere.\cite{elasticity}

The presence of this singular spectrum suggests that 
microwave conductivity experiments 
\cite{lewis04,pinczuk} may be able to discern such structures as a 
signature of, e.g., the quantum Hall nematic suggested by
magnetotransport experiments.\cite{lilly99andothers99,competition}

\begin{figure}
\begin{center}
\leavevmode
\includegraphics[width=2.6in]{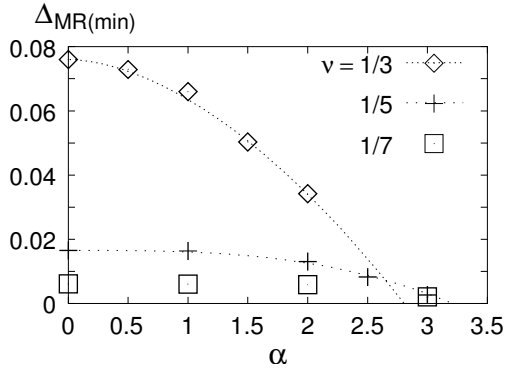}
\vspace{-.5cm}
\end{center}
\caption{\label{fig:collapse} 
Possible collapse of the magnetoroton minimum as function of the
anisotropy parameter $\alpha$.  Dotted lines are guides to the eye.  
} 
\vspace{-.5cm}
\end{figure}

Another interesting feature (Fig.\ \ref{fig:collapse}) is that the
magnetoroton minima and the gap at the origin appears to collapse at
some finite anisotropy factor $\alpha$.  
The appearance of elastic modes that go soft may be a precursor 
of the appearance of charge density waves in the system.\cite{liqcryst04}
Spectra in higher LL's can be obtained by multiplying the projected
density operator $\bar{\rho}_{\bf q}$ by $L_L(q^2/2)$, where $L_L(x)$ is
the Laguerre polynomial and $L$ corresponds to the desired
LL.\cite{higherLLsVq}  The rightmost panel of Fig.\ \ref{fig:spectra}
shows the modification of the SMA spectrum for the first excited LL.
Numerical error due to the large $L_L(q^2/2)$ factors makes it
difficult to get reliable results for higher LL's at this point.

Theoretical predictions of which type of order is lowest in {\em free}
energy (required to decide which state is favorable at {\em finite
temperatures}, as shown in Refs. [\onlinecite{competition,melting17}])
are still incomplete, however, as the entropy is likely to be
dominated by gapless modes originating, e.g., from the Goldstone modes
associated with the spontaneous breaking of the continuous rotational
symmetry of the isotropic states.  Generalization of the SMA to
gapless rotational modes (see e.g. \onlinecite{radzihovsky02fogler04})
will require, however, 3-body operators which demanding considerably
higher computing capabilities.\cite{foglerprivate}

\section{Summary}  

Summarizing: in connection with the recent variety of experimental
evidence supporting liquid crystalline phases in quantum
Hall systems,\cite{lilly99andothers99,competition,melting17,lewis04}
we have calculated the collective excitation spectrum in
the SMA approximation for a quantum Hall liquid crystal.  
We found that the spectrum of excitations coupled to the ground state
by the density operator remains gapped, but develops a significant
anisotropy, which in the case of the nematic and tetratic liquid
crystals has a singular gap in the long wavelength limit.

\section{Acknowledgments}

We would like to acknowledge helpful discussions with G.\
Vignale, O.\ Ciftja, A.T.\ Dorsey, M.\ Fogler, A.\ MacDonald, S.\
Girvin, E.\ Fradkin, H.\ Fertig, and J.\ Jain.
Acknowledgment is made to the University of Missouri Research Board
and Research Council, and to the Donors of the Petroleum Research
Fund, administered by the American Chemical Society, for support of
this research.


\end{document}